\documentclass[fleqn,10pt]{wlscirep}
\usepackage[utf8]{inputenc}
\usepackage[T1]{fontenc}
\usepackage{lineno}
\usepackage{booktabs}
\usepackage{multirow}
\usepackage{graphicx}
\usepackage{array}
\usepackage{listings}

\title{Machine learning-based correlation analysis of decadal cyclone intensity   with sea surface temperature: data and tutorial }

\author[1,2]{Jingyang Wu} 
\author[2,*]{Rohitash Chandra}
\affil[1]{Transitional Artificial Intelligence Research Group, School of Mathematics and Statistics, UNSW Sydney, Australia}
\affil[2]{Centre for Artificial Intelligence and Innovation, Pingla Institute, Sydney, Australia}

\affil[*]{corresponding author: Rohitash Chandra (rohitash.chandra@unsw.edu.au)}

\affil[$\dag$]{these authors contributed equally to this work}

\begin{abstract}  
The rising number of extreme climate events in the past decades has motivated the need for a thorough consideration of tropical cyclone genesis and intensity, given the sea-surface temperature (SST). In this paper, we present an analysis of the relationship between the increasing global SST with cyclone genesis using linear regression machine learning models. We extract and curate a dataset of tropical cyclones across selected ocean basins with their associated  SST over the past 40 years. We provide correlation analysis using linear regression and visualisation strategies. Our preliminary results show a strong positive correlation between SST and high wind speed   (Category 5 Cyclones) across the South Indian Ocean and the North Indian Ocean via linear regression and machine learning models.  Our dataset and available open-source code offer a novel perspective for the investigation of the genesis and intensity of tropical cyclones. Alongside the time and position of each cyclone, we also provide the related Saffir-Simpson category, season, wind speed, and SST for 15 days before and after the tropical cyclone genesis.

\end{abstract}
\begin{document}


\flushbottom
\maketitle

\thispagestyle{empty}


\section{Introduction}

Tropical cyclones (hurricanes and typhoons) are extreme meteorological events that possess enormous force and cause extensive destruction. High-intensity tropical cyclones can result in significant consequences, such as casualties, economic losses, infrastructure destruction, and flooding, among other problems \cite{needham2015review,lenzen2019economic,woodruff2013coastal}. It has been estimated that tropical cyclones led to 2.6 million deaths globally in the last two centuries, with an annual average of 13,000 deaths \cite{nicholls2003expert}. The destructive power of tropical cyclones is related to their sustained and intensive wind speed, heavy rainfall and associated floods. According to the Saffir-Simpson Hurricane Wind Scale (SSHWS) \cite{taylor2010saffir}, the maximum wind speed (wind intensity) at the centre of a tropical cyclone can exceed 252 km/h. In general, the destructive force of tropical cyclones multiplies by around four for each level rise on the SSHWS \cite{pielke2008normalized}. Extremely strong wind speed can destroy civil infrastructure systems, resulting in loss of life \cite{loggins2015rapid}. Furthermore, tropical cyclones also incur additional hazards such as landslides and flooding events resulting from heavy rainfall, with tornados and wave surges  \cite{taylor2010saffir}. The main areas affected by tropical cyclones include southern Asia (India, the Philippines, etc.), the Southeastern region of the United States (Florida, Louisiana), and Northern Oceania (Australia) \cite{sen2023impact,chikodzi2021linking}. Therefore, accurate cyclone intensity prediction is essential for minimising damage and disaster management. Forecasting the early stages enables prompt evacuations, improved readiness, and the execution of preventive measures to mitigate the severe impact on communities and infrastructure. 


Global warming has been a major concern in recent decades \cite{botkin2007forecasting} and cyclones develop owing to conditions such as ocean temperature, humidity, and atmospheric pressure \cite{needham2015review}. Based on historical data and observations, cyclones frequently occur during the highest temperatures of the year's hottest months in the specified region \cite{emanuel2003tropical}. The correlation between the continuing rise in sea surface temperature and the occurrence of cyclones has been established. Another climate phenomenon that supports global warming is called El Niño, which is defined as a 5-month moving average of Sea Surface Temperature (SST) anomalies in the equatorial region of the Pacific Ocean exceeding 0.4°Celcius for 6 months or more \cite{trenberth1997definition}. Some system models have confirmed that the rise in SST during El Niño events leads to an increase in extreme climate \cite{carreric2020change}. The available evidence strongly indicates that alterations in SST are associated with tropical cyclones, potentially influencing their formation, intensity and duration. Some researchers utilise the North Atlantic Stochastic Hurricane Model, in combination with relative-SST calculated by differences in local SST and the global tropical SST, to estimate regional changes in the United States cyclone activity for the 2030s \cite{hall2021us}. Another investigation \cite{bui2021influence} indicates that SST is strongly correlated with sensible and latent heat fluxes and that these factors influence the strength and frequency of extratropical cyclones. However, barely sufficient evidence establishes a direct correlation between tropical cyclones and SST. A comprehensive dataset recording multiple characteristics of tropical cyclones and corresponding SST data will help researchers examine the relationship between these features.

A General Circulation Model (GCM) \cite{phillips1956general,
roeckner1996atmospheric,wilby1997downscaling,weart2010development} is a sophisticated climate model used to simulate the Earth's atmosphere, oceans, and land surface processes. GCMs use mathematical equations to represent physical processes such as radiation, convection, and the water cycle. Phillips \cite{phillips1956general}  developed the first GCM  by enhancing an existing climate model that showed how energy and momentum are transferred in giant eddies in the atmosphere \cite{weart2010development}. Due to advancements in computer processing capabilities, GCMs have incorporated additional physical equations to improve modelling accuracy. These equations now consider a wider range of climate conditions and geographical locations \cite{wilby1997downscaling,grotch1991use}. Recent work has incorporated tides and internal gravity waves into  GCM designs, hence enhancing the realism of the simulated circulation system\cite{arbic2010concurrent,arbic2012global,arbic2018primer}. Certain works propose the utilisation of intricate physical and mathematical equations to enhance the simulation efficacy of GCM, including stochastic dynamic methods, semi-implicit time integration algorithms, and the splitting of k-turbulence equations \cite{galin2019dynamic,kulyamin2014atmospheric,moshonkin2018algorithm}. At present, GCM is considered a dependable method for acquiring temperature data and a source for sea surface temperature data.


Although there have been advances in meteorological modelling, forecasting the pathway and intensity of tropical cyclones remains a significant challenge, particularly within the rapidly changing global climate. While existing studies acknowledge the influence of  SST and climatic phenomena such as El Niño on the dynamics of the cyclones, significant gaps persist in quantifying these relationships and translating them into actionable forecasting frameworks \cite{roy2012tropical}. We aim to create a comprehensive dataset that integrates information on tropical cyclones and corresponding SST data, utilising modern data analytics and machine learning techniques to dive into the nuanced relationships between SST fluctuations and cyclone behaviour. This dataset is designed to investigate how global temperature variations influence tropical cyclone activity, ultimately providing a robust theoretical foundation for improving prediction models.

In this study, we extract and curate a dataset of tropical cyclones across selected ocean basins with their associated  SST over the past 40 years. Our dataset and available code software offer a novel perspective for the investigation of the genesis and intensity of tropical cyclones. Alongside the time and position of each cyclone, we also provide the related SSHWS, season, wind speed, and SST for 15 days before and after the tropical cyclone genesis. We provide analysis using linear regression and machine learning models to determine the correlation between cyclone genesis, cyclone intensity category, and SST.

\section{Methods}




\subsection{Cyclone-SST Dataset}

The Cyclone-SST dataset is derived from two primary sources. The Joint Typhoon Warning Centre (JTWC) acquires and monitors tropical cyclones using several technologies, including satellites, radars, buoys, and scatterometers. Presently, numerous studies utilise JTWC's tropical cyclone data for analysis and comparison \cite{yurchak2018estimation,demaria2014tropical,huang2021evaluation}, which serves as a dependable source for cyclone information. We combine GCM-based SST data from National Oceanic and Atmospheric Administration (NOAA) with tropical cyclone track records from JTWC to create a new dataset. Our dataset includes data for cyclone tracks in selected basins since 1981, as well as SST records for several days preceding and following the tropical cyclones. In addition, the dataset contains SSHWS, seasonal information, intensity of tropical cyclones etc. This generates a dataset that can be comprehensively analysed by others. Both sources of data are described separately in the following subsections.

\subsection{NOAA SST data}

\begin{figure}[ht]
    \centering
    \includegraphics[width=\linewidth]{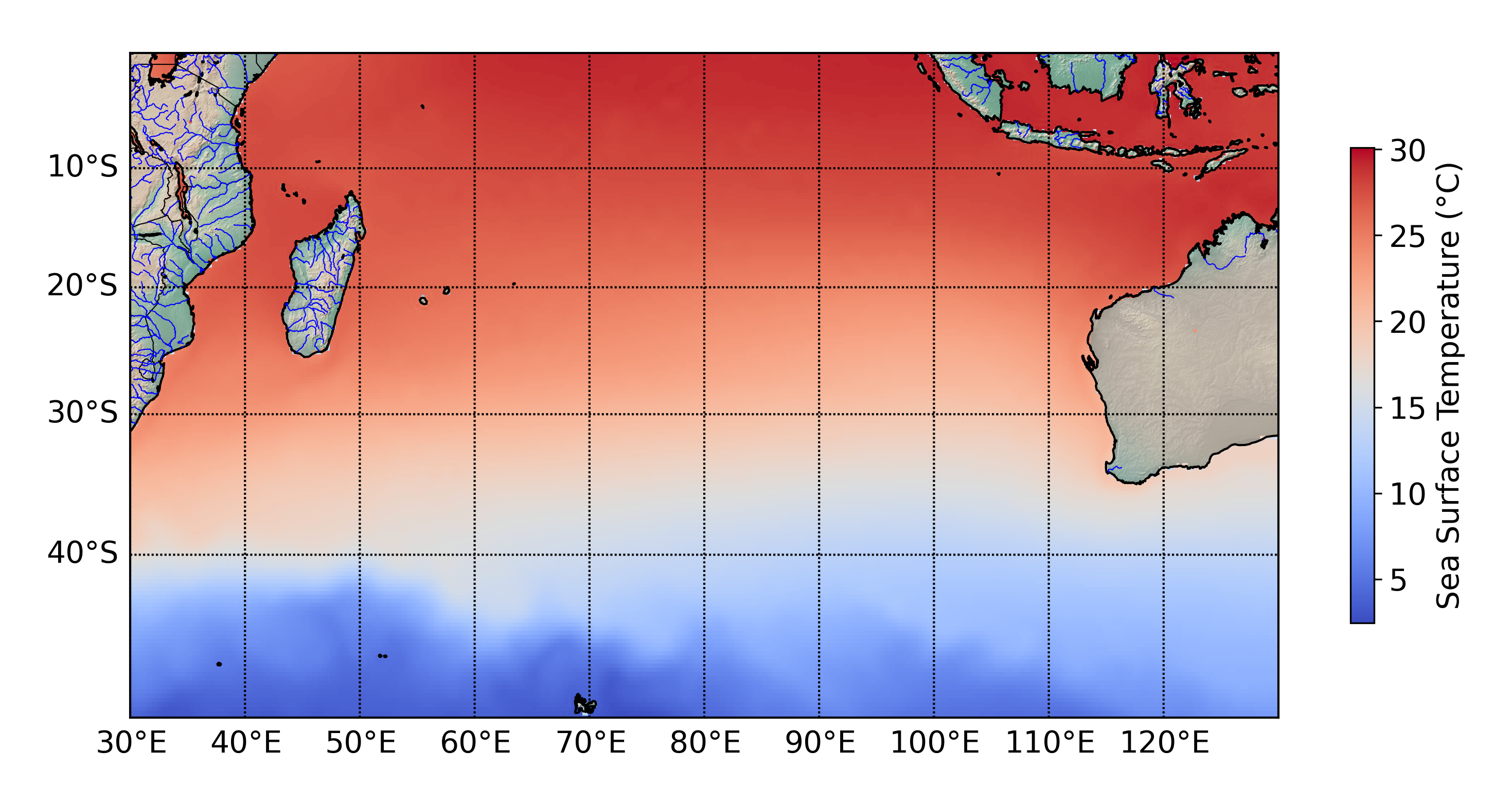}
    \caption{ The heat-map of average  SST  from 1981 to 2024, derived from OI-SST, using the southern Indian Ocean as an example (0°S to 50°S, 30°E to 135°E). The temperature decreases with increasing latitude, and the legend indicates that the temperature transitions from low to high.}
    \label{fig_si_avgsst}
\end{figure}

The NOAA 1/4° Daily Optimum Interpolation SST \cite{huang2021improvements} (OI-SST) is a long-term Climate Data Record that incorporates observations from different platforms (satellites, ships, buoys and Argo floats) into a regular global grid \footnote[1]{\url{https://www.ncei.noaa.gov/products/optimum-interpolation-sst}}. The dataset is interpolated to fill gaps on the grid and create a spatially complete map of SST. Satellite and ship observations are referenced to the buoy to compensate for platform differences and sensor biases. The data is stored in a netCDF4 file, consisting of one variable denoting SST and four dimension indicators: longitude, latitude, time, and experimental version. The dataset collects monthly SSTs from various basin regions, covering the period from 1979 to 2021 and documented in datetime64[ns] format. On OI-SST, the Earth is divided into a 0.25 × 0.25 degree grid, with each grid allocated the monthly average SST measured in Kelvin's heat units. Figure \ref{fig_si_avgsst} shows the mean SST between 1971 and 2021 extracted from OI-SST, taking the southern Indian Ocean as an example (0°South to 50°South, 30°Eeast to 135°Eeast). The legend on the right illustrates the temperature scale, with colors ranging from blue to red representing increasing temperatures from 0°C to 30°C. Regions closer to the equator receive more direct and consistent solar radiation throughout the year, leading to greater heat accumulation at the ocean surface. This pattern is evident in the figure, where regions closer to the equator are shown in red, indicating higher temperatures, while areas farther from the equator appear in blue, reflecting cooler conditions.

\begin{figure}[ht]
    \centering
    \includegraphics[width=\linewidth]{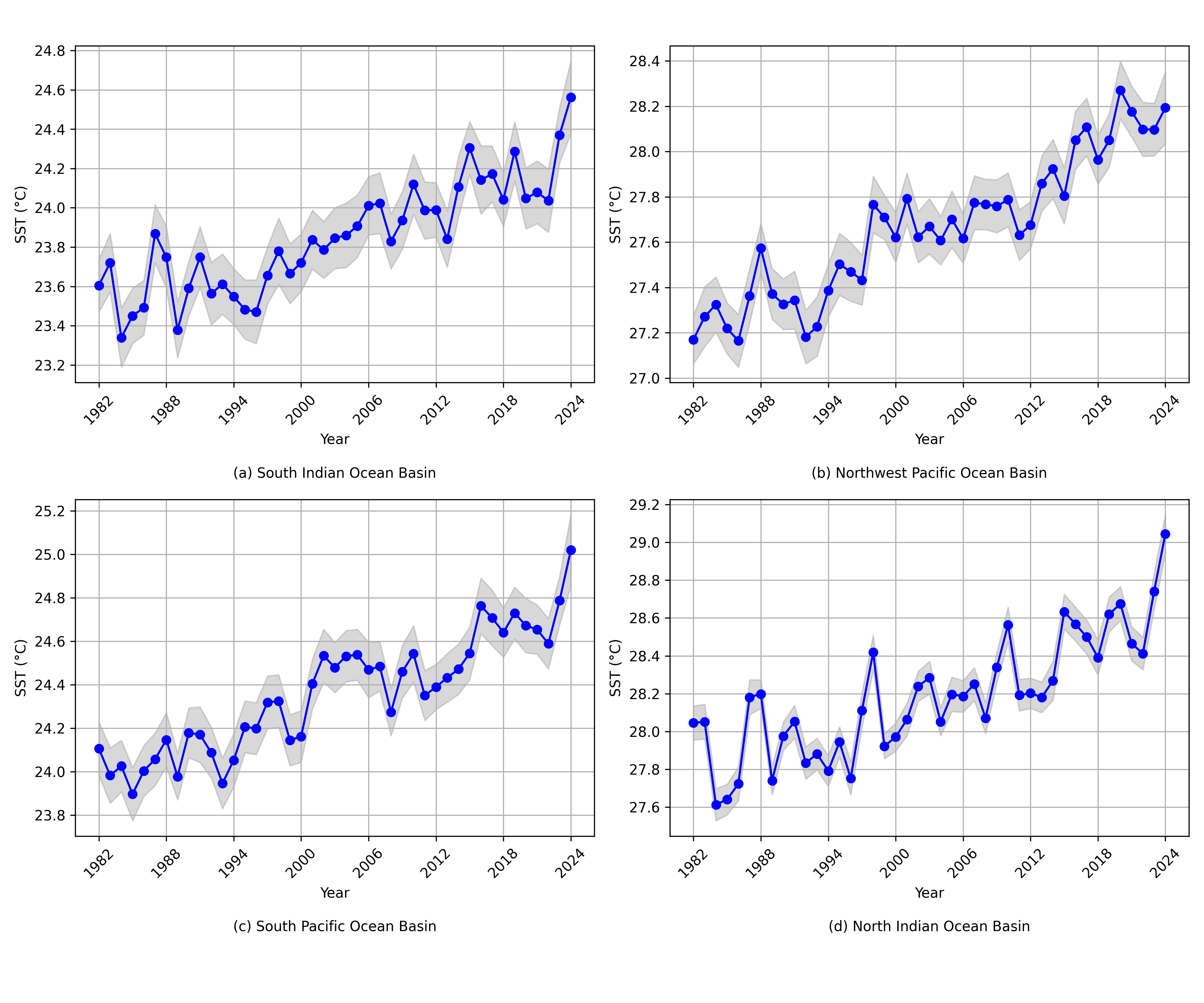}
    \caption{The annual mean SST at the cyclone centre points from 1981 to 2024: (a) South Indian Ocean Basin (b) Northwest Pacific Ocean Basin, (c) South Pacific Ocean Basin,  (d) North Indian Ocean Basin. Thus, the blue line indicates the annual mean, while the grey shading represents the 95\% confidence interval.} 
    \label{fig_yearsst}
\end{figure}

 Figure \ref{fig_yearsst} presents the annual mean SST and its associated standard deviation across four different basins. Although the SST in the four basins exhibited fluctuations from 1982 to 2024, an overall increase was observed over these decades. The standard deviation of SST within the four basins remains relatively stable, with the highest values observed in the South Pacific Ocean region. In contrast, the North Indian Ocean region shows the lowest standard deviation. 

\subsection{Tropical Cyclone  Data}

\begin{figure}[ht]
    \centering
    \includegraphics[width=\linewidth]{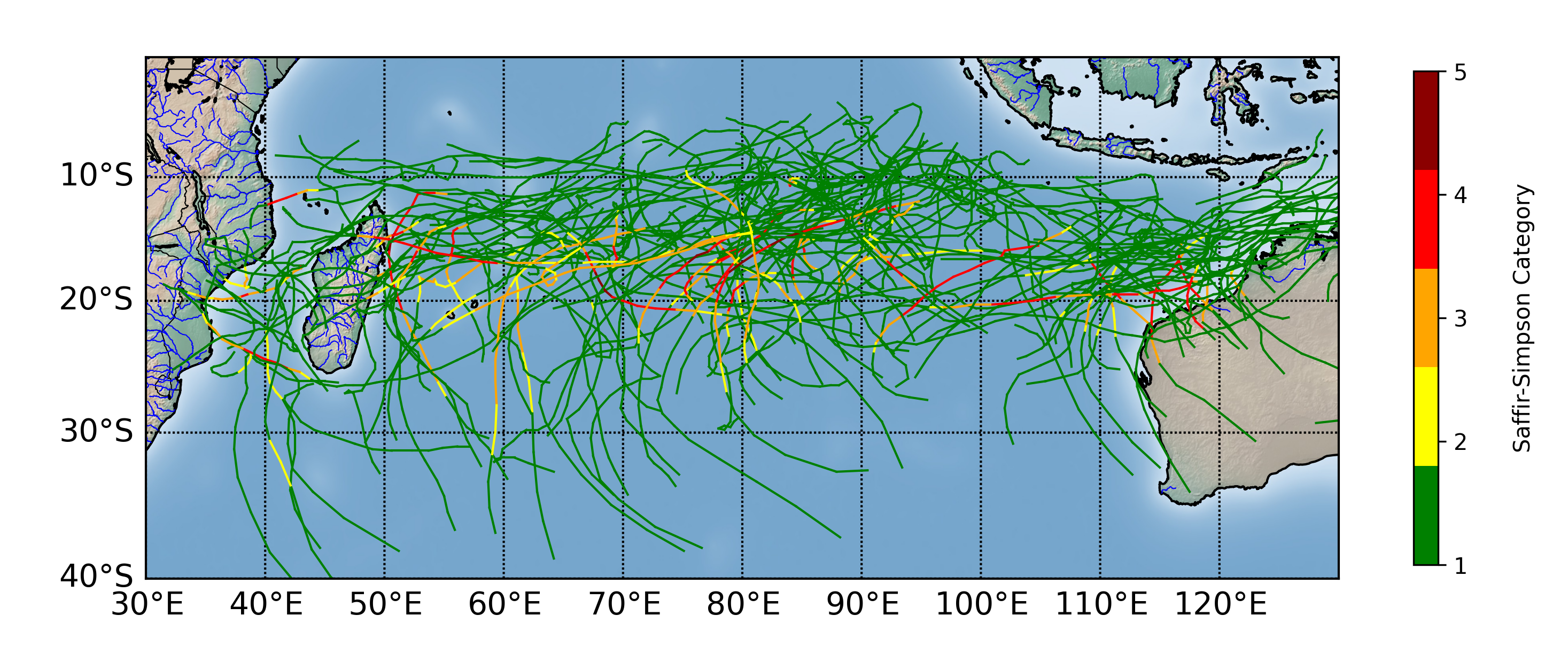}
    \caption{ The diagram illustrates the trajectories of the tropical cyclones, represented by a series of data points from 1981 to 2024. We selected the tropical cyclones randomly to serve as examples using the South Indian Ocean basin as the region.
    We colored each trajectory of the cyclones based on the maximum Saffir-Simpson wind category reached during the cyclone's existence. Specifically, green, yellow, orange, red, and dark red correspond to Category 1 through Category 5 cyclones, respectively.}

    \label{fig_si_cyctracks}
\end{figure}

We source tropical cyclone data from the Joint Typhoon Warning Centre of the US Navy \footnote[2]{\url{https://www.metoc.navy.mil/jtwc/jtwc.html?best-tracks}}. The JTWC preserves an archive of TC track data, typically known as "best-tracks". Each best-track file comprises tropical cyclone centre coordinates and intensities (i.e., the maximum 1-minute mean sustained wind speed at 10 meters) recorded at six-hour intervals. The geographical domain of the archive is the western North Pacific (WP), North Indian Ocean (IO) and Southern Hemisphere (SH). We merged the cyclone temporal data for each basin into a single CSV file. Figure \ref{fig_si_cyctracks} displays the movements of cyclones included in our dataset. A random selection of 30\% of the cyclones from the South Indian Basin was plotted on the corresponding shaded maps. The figure indicates that cyclones with intensity achieving higher SSHWSs predominantly form within the latitude range of 10° to 20° south.

\begin{figure}[ht]
    \centering
    \includegraphics[width=\linewidth]{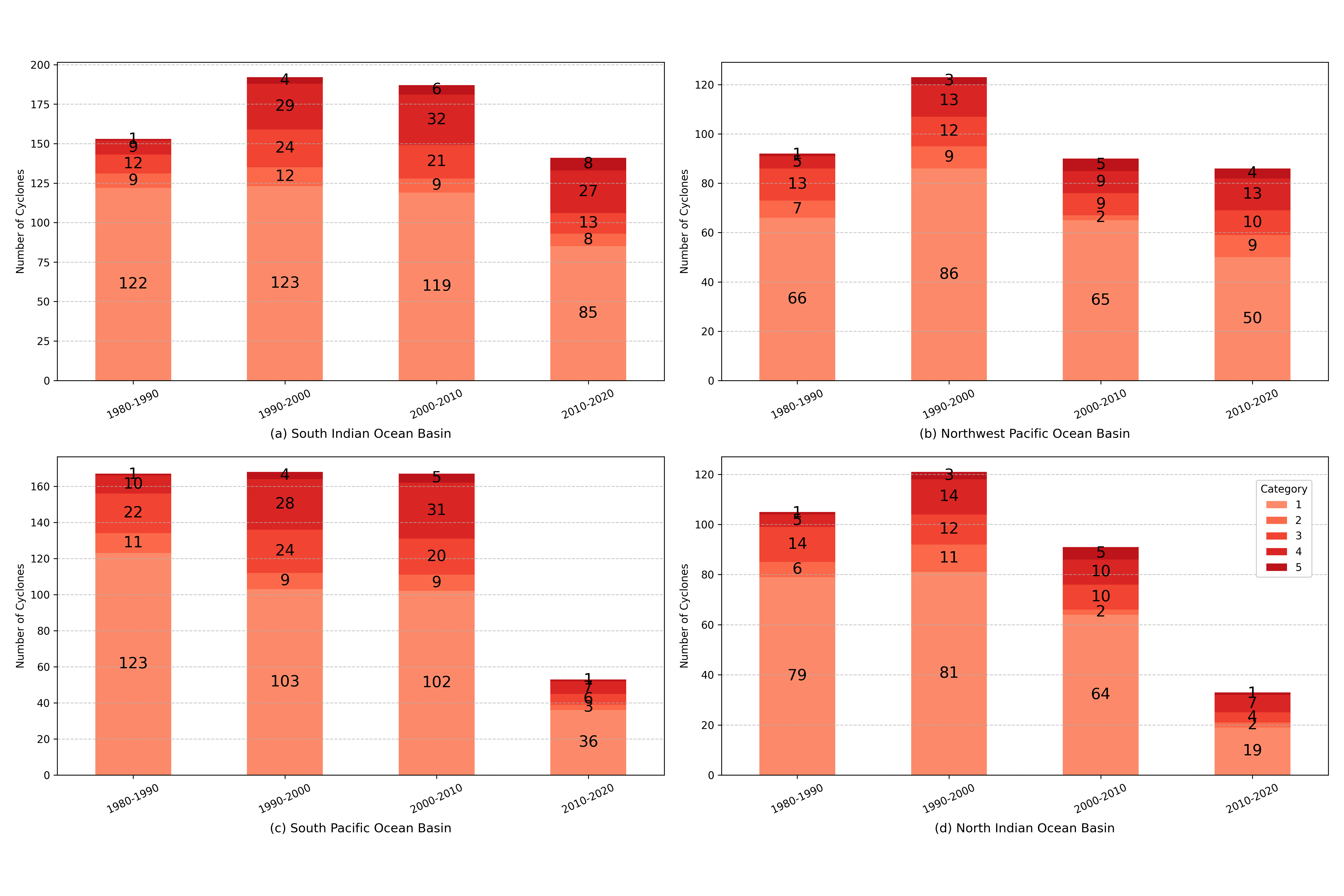}
    \caption{The decadal frequency of tropical cyclones across different ocean basins, categorised by their peak wind intensity:(a) South Indian Ocean Basin (b) Northwest Pacific Ocean Basin (c) South Pacific Ocean Basin (d) North Indian Ocean Basin. 
    The colour gradient transitions from light for category 1 to dark for category 5 based on Saffir-Simpson scale.}
    \label{fig_decade_cyclone_counts}
\end{figure}

We provide a clearer depiction of the number of cyclones at various wind intensity levels across 4 different basins,   with their changes over the past 4 decades, with statistical analysis in Figure \ref{fig_decade_cyclone_counts}. We can observe that apart from the basins lacking cyclone records after 2010, the overall proportion of cyclones reaching higher wind intensities has risen, indicating that extreme weather events have become more frequent over time.

\subsection{Data curation framework}

\begin{figure}[htbp!]
    \centering
    \includegraphics[width=\linewidth]{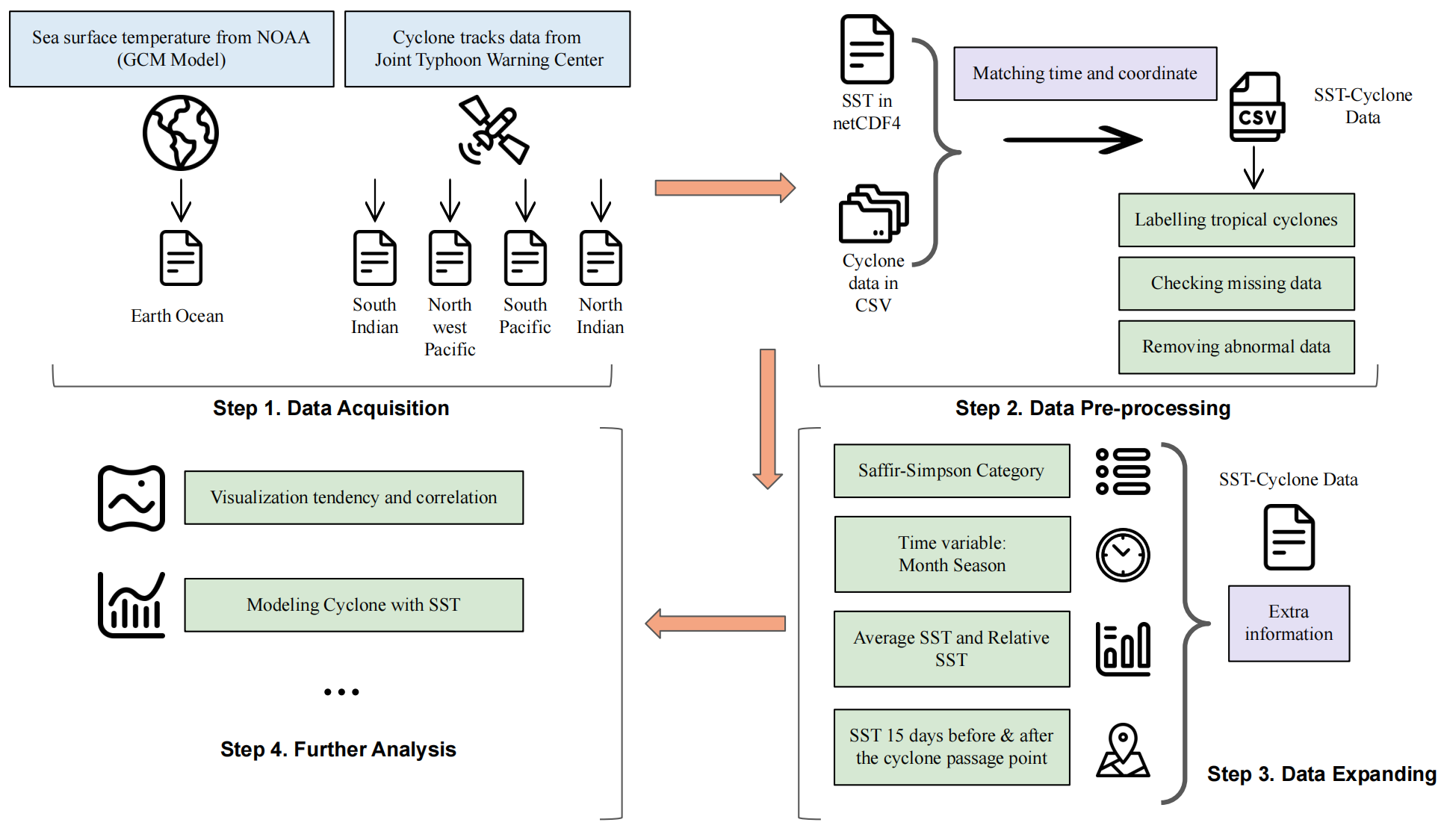}
    \caption{Framework illustrating the process of acquiring SST data from NOAA’s GCM model and cyclone track data from the JTWC. We merge them based on time and location, preprocessing, expanding the dataset (e.g., adding Saffir–Simpson categories and relative SST), and finally conducting further analyses such as visualisation, correlation studies, and modelling. }
    \label{fig_framework}
\end{figure}

This framework (Figure\ref{fig_framework}) produces a dataset that combines SST with tropical cyclone track and wind intensity. The first step involves collecting SST data from NOAA. This data is obtained through GCMs, which simulate interactions between the Earth's atmosphere, oceans, and lithosphere.

In the second stage, tropical cyclone data is obtained from the JTWC by utilising estimations derived from satellite observations. This data offers comprehensive information on the trajectory, intensity, and characteristics of tropical cyclones, including their geographic coordinates and period, which are crucial for integrating with SST data. The third step involves pre-processing the cyclone datasets. We combined all the text files acquired from JTWC into a data file and arranged all tropical cyclones in numerical sequence. We removed tropical cyclones with missing information from the dataset and also provided  details for the cyclones that have been removed. The pre-processing stage ensures the data is clean and standardised for subsequent analysis. In the final stage, we match the SST with the geographic location and time of each tropical cyclone, and the SST information will be added to the data file of tropical cyclones. In addition, we generate a set of supplementary columns depicting the season of tropical cyclones, the Saffir-Simpson category, and the SST throughout various times of the tropical cyclones. This step results in a refined dataset that can be used for analysing the relationship between SST and tropical cyclone characteristics.

\subsection{Data Records}

We sourced the raw data files  from the NOAA and JTWC websites, which are available in our Github repository \footnote{\url{}}. For the tropical cyclone dataset in each basin, we provide a merged data file (CSV) in GitHub labelled "Basin\_hurricane". For SST data, we provide code to download and merge these NetCDF files. We can utilise the Python file labelled "data\_merge" to merge the final dataset. The final dataset is stored as CSV files, with one file containing cyclone information from different basins. The datasets from different basins are identified by their file names; for instance, the South Indian Basin's data is labelled "South\_indian\_hurricane\_with\_SST \footnote{Githubrepo link for the files: \url{....}}."



\section{Results}

\subsection{Cyclone charateristics}

We present two summary tables to provide an overview of key characteristics within the cyclone datasets. Table \ref{tab_1} reports the frequency and intensity of cyclones in the four basins, including the number and proportion of cyclones that reach the Saffir-Simpson Hurricane Wind Scale (SSHWS) at levels 3, 4, and 5. A total of 2030 cyclones are included across all the regions, with the South Indian Ocean contributing the largest share (670 events), followed by the North Indian Ocean (564), the Northwest Pacific (405), and the South Pacific (391). Tropical cyclones SSHWS 3 or higher are typically classified as intense cyclones. In terms of proportional occurrence, SSHWS 3 storms are most frequent in the North Indian Ocean, SSHWS 4 storms peak in the South Indian Ocean, while the South Pacific basin exhibits the highest frequency of SSHWS 5 cyclones. 

Table \ref{tab_2} provides a statistical overview of  SST  characteristics associated with cyclones across basins, including values from 15 days before to 15 days after each event. On average, SSTs are higher in the South Indian and South Pacific basins, with values exceeding 27°C throughout the observation window. In contrast, the Northwest Pacific consistently shows the lowest SST values, with a mean SST below 26°C at all time points. We can observe that the SSTs peak in the days preceding cyclone passage and tend to decline after the event, particularly in the Indian Ocean basins. For instance, SST in the South Indian Ocean drops from 27.72°C at 5 days before the cyclone to 27.14°C at 5 days after the cyclone, reflecting post-storm ocean cooling. The mean SSHWS also varies by region, with the Northwest Pacific exhibiting the highest average intensity (2.36), and the South Pacific the lowest (1.73). Relative SST values (defined as anomalies from regional means) are near zero across all basins, indicating that storm genesis tends to occur under typical background thermal conditions rather than during exceptional warming periods. We provide a more detailed analysis of the impact of cyclones on sea surface temperature in the following subsections. 

\begin{table}[htbp!]
\centering 
\begin{tabular}{|l|l|p{4cm}|p{4cm}|p{4cm}|}
\hline 
Basin & Count & Number of cyclones reaching Saffir-Simpson Category 3 & Number of cyclones reaching Saffir-Simpson Category 4 & Number of cyclones reaching Saffir-Simpson Category 5\\
\hline
South Indian & 670 & 183 (27.31\%) & 115 (17.16\%) & 19 (2.84\%)\\
\hline
South Pacific & 391 & 44 (11.25\%) & 40 (10.23\%) & 13 (3.32\%)\\
\hline
North Indian & 564 & 159 (28.19\%)  & 86 (15.25\%) & 11 (1.95\%)\\
\hline
NorthWest Pacific & 405 & 92 (22.72\%) & 48( 11.85\%) & 10 (2.47\%)\\
\hline

\end{tabular}

\caption{Statistical summary of cyclone frequency and intensity across four basins}
\label{tab_1}
\end{table}

\begin{table}[htbp!]
\small
  \centering 
  \begin{tabular}{|p{4.5cm}|c|c|c|c|c|c|c|c|}
    \hline 
    \textbf{Variable} 
    & \multicolumn{2}{c|}{\textbf{South Indian}} 
    & \multicolumn{2}{c|}{\textbf{South Pacific}} 
    & \multicolumn{2}{c|}{\textbf{North Indian}} 
    & \multicolumn{2}{c|}{\textbf{NorthWest Pacific}} \\
    \hline 
    & Mean & Std & Mean & Std & Mean & Std & Mean & Std \\
    \hline
    SST -15 days (°C)   & 27.64 & 1.73 & 28.00 & 1.88 & 26.87 & 2.04 & 25.98 & 2.56 \\
    \hline
    SST -10 days (°C)   & 27.70 & 1.70 & 28.03 & 1.86 & 26.89 & 2.05 & 25.98 & 2.56 \\
    \hline
    SST -5 days (°C)    & 27.72 & 1.65 & 28.02 & 1.83 & 26.89 & 2.05 & 25.93 & 2.59 \\
    \hline
    SST on event (°C)   & 27.44 & 1.74 & 27.81 & 1.89 & 26.87 & 2.04 & 25.87 & 2.59 \\
    \hline
    SST +5 days (°C)    & 27.14 & 1.56 & 27.56 & 1.79 & 26.91 & 2.07 & 25.86 & 2.60 \\
    \hline
    SST +10 days (°C)   & 27.36 & 1.55 & 27.63 & 1.77 & 26.96 & 2.09 & 25.85 & 2.62 \\
    \hline
    SST +15 days (°C)   & 27.52 & 1.54 & 27.73 & 1.81 & 27.01 & 2.09 & 25.84 & 2.62 \\
    \hline
    
    Relative-SST   & -0.07 & 0.65 & -0.03 & 0.61 & -0.07 & 0.53 & -0.03 & 0.43 \\
    \hline
    Saffir-Simpson Scale   & 1.81 & 1.25 & 1.73 & 1.20 & 1.79 & 1.21 & 2.36 & 1.71 \\
    \hline
    \hline
  \end{tabular}
  
    \caption{Summary of cyclone variables (SST) across four basins, including mean and standard deviation}
    \label{tab_2}
\end{table}


\subsection{Correlation analysis between tropical cyclones and SST}

\begin{figure}[ht]
\centering
\includegraphics[width=\linewidth]{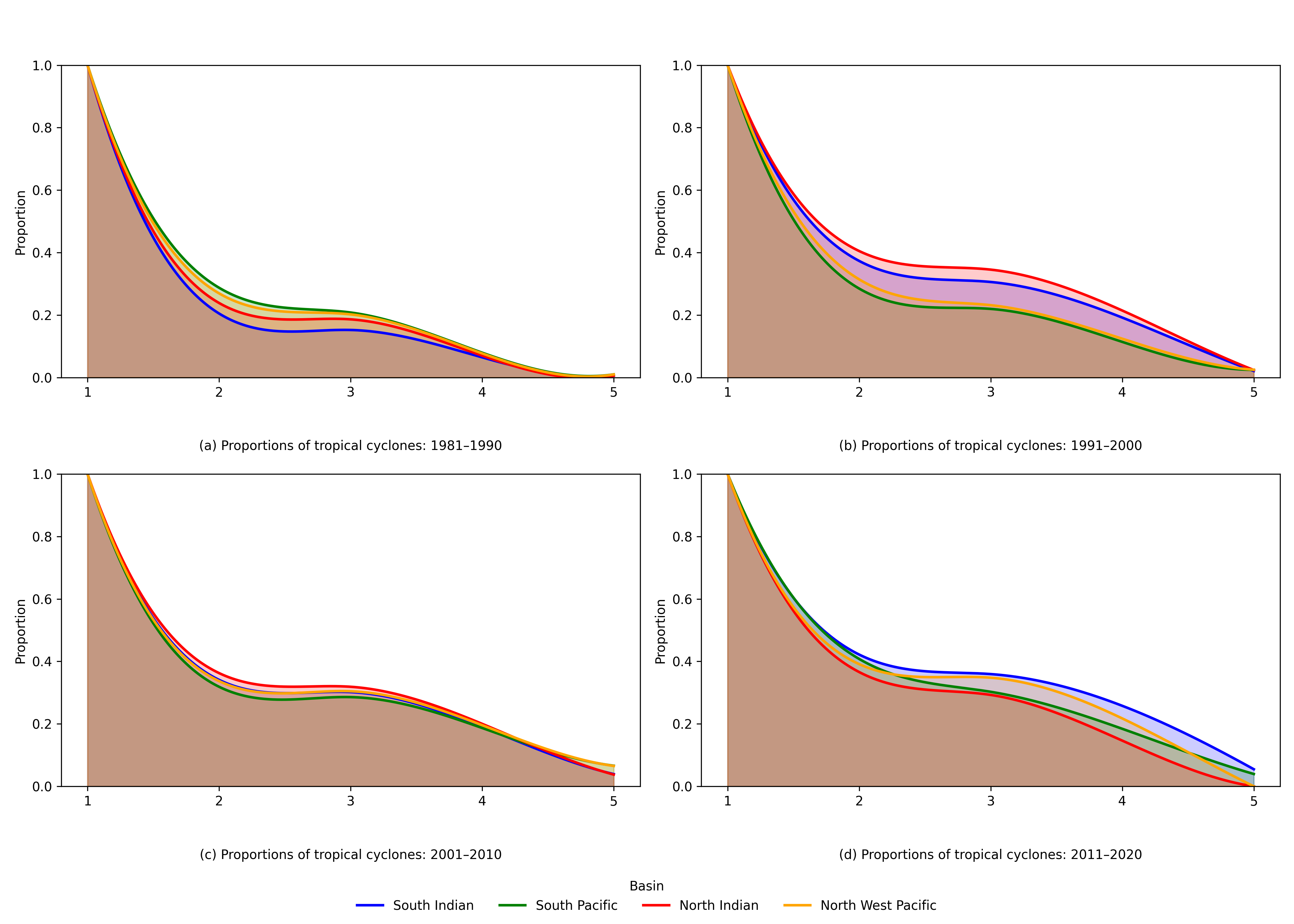}
\caption{The proportions of tropical cyclones of varying intensities within each decade, expressed as percentages of the total number of cyclones. The lines and shading in different colours indicate data from distinct ocean basins. Subplots (a), (b), (c), and (d) correspond respectively to the four decade periods between 1980 and 2020.}
\label{fig1}
\end{figure}

 We performed correlation analysis and visualisation of the data to verify the reliability and usability of our dataset. First, we visualised the percentage of tropical cyclones of different strengths in different regions in each decade since 1980. This can be seen as an extension of the cyclone statistics in Table \ref{tab_1}. We employ the  SSHWS in the dataset as a criterion for assessing tropical cyclones. In Figure \ref{fig1}, the cyclones display a pattern of growing strengthening followed by a decline in wind intensity. A large number of cyclones reached their peak Saffir-Simpson level at 1, as our analysis indicates a significant decrease in the ratio of Category 1 to Category 2 tropical cyclones across any period. The diagram also indicates that throughout the two-decade intervals of 1990 to 2000 and 2010 to 2020, there was an increase in tropical cyclones exhibiting high wind intensity.

According to  Dare et al. \cite{dare2011threshold}, before a tropical cyclone is formed, the SST is generally stable or may increase slightly, especially in tropical and subtropical regions, where higher SSTs are conducive to the formation and intensification of tropical cyclones. During a tropical cyclone, strong winds and storm activity mix surface water with cooler, deeper water in a process known as ocean mixing, causing a significant drop in SST. After the tropical cyclone, the SST begins to slowly recover, but the recovery rate depends on factors such as ocean currents, wind patterns, and regional climatic conditions.

Therefore, we offer a methodology to examine the temperature changes of the sea surface before and after the cyclones. The Generalised Additive Model (GAM) \cite{hastie1986generalized}, a regression model that is capable of handling nonlinear trends in data and been prominent in ecology and environmental modelling \cite{
yee1991generalized,ravindra2019generalized}. We employ GAMs to fit (model) the temperature changes of the SST for 15 days prior to and after a cyclone. 

\begin{figure}[htbp!]
\centering
\includegraphics[width=\linewidth]{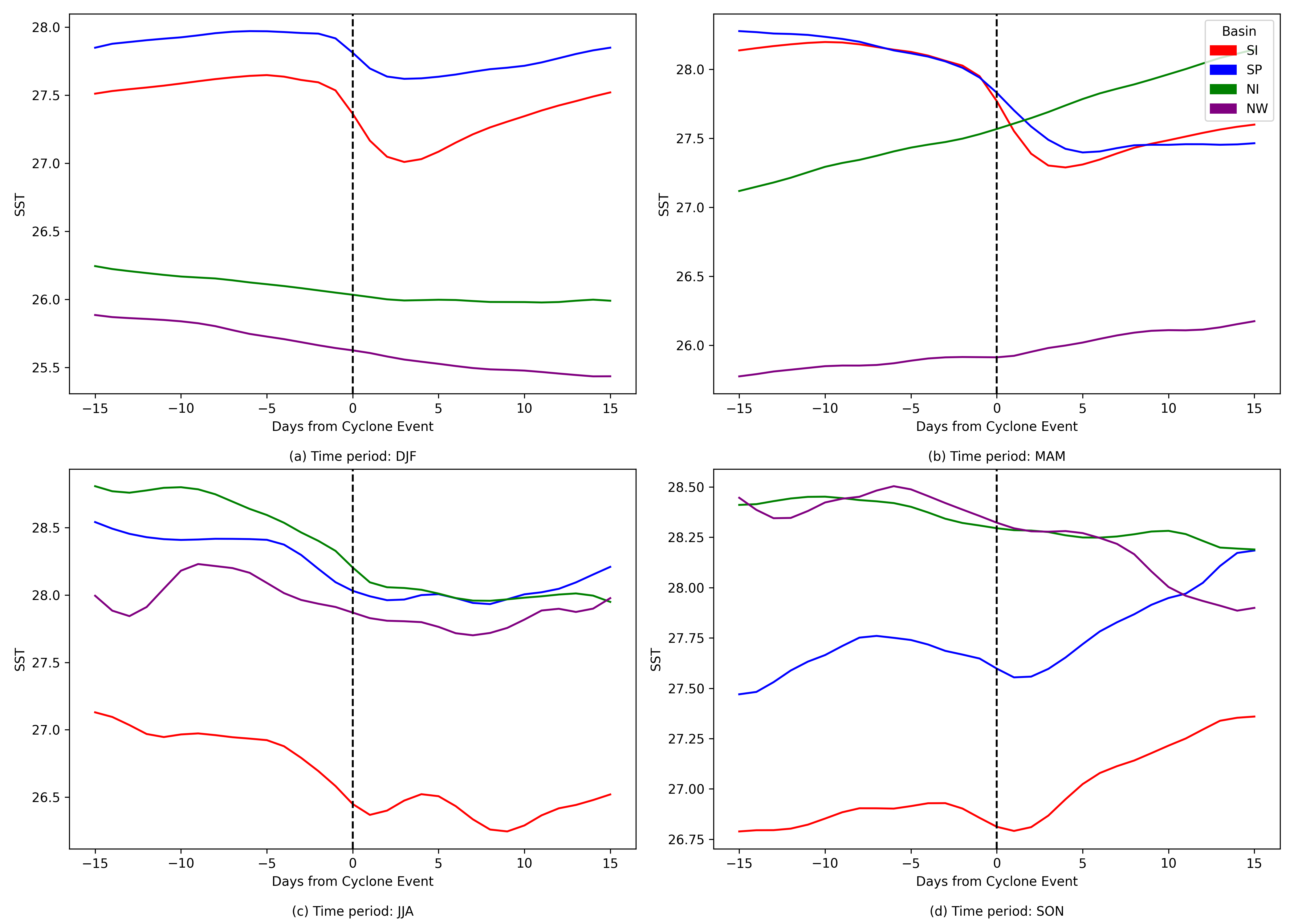}
\caption{SST change over 15 days before and after cyclone events across different basins (SP, SI, NI, NWP), as modeled by the GAM. 
}
\label{fig2}
\end{figure}

Figure \ref{fig2} presents the results of the GAM, with the four sub-figures depicting the results over the different seasons (DJF,MAM,JJA,SON). We can observe a sharp SST drop in the South Indian Ocean (SI) during DJF (Panel a) immediately following the cyclone event, with the minimum occurring around day +3. We observe a similar, albeit more gradual, cooling trend  in the North Indian Ocean (NI). In contrast, the Northwest Pacific (NW) and South Pacific (SP) basins exhibit relatively stable SST patterns. We find that average SSTs in the South Indian Ocean (SI) and South Pacific Ocean (SP) were higher than those in the North Indian Ocean (NI) and Northwest Pacific Ocean (NW), consistent with the general seasonal pattern of elevated temperatures in the Southern Hemisphere during the DJF period.

 During MAM (Figure \ref{fig2}  Panel b), the SST response is particularly strong in both the South Indian (SI) and North Indian (NI) basins. We also observe that the SST declines rapidly following cyclone passage, with decreases exceeding 1°C within five days. The South Pacific (SP) basin shows a different pattern: a slight initial cooling is followed by a gradual convergence with SI and NI temperatures after day +5. This temporal alignment suggests a coherent upper-ocean adjustment across these regions, likely associated with enhanced vertical mixing and upwelling induced by cyclone activity. In contrast, the Northwest Pacific (NWP) basin exhibits a continuing warming trend, which may reflect its subtropical location and weaker mixing response during this season.

 The JJA (Figure \ref{fig2}  Panel c) results indicate that SI undergoes the most intense cooling, reaching its lowest SST (~26.3°C) around day +7, whereas other basins show relatively flatter responses. Furthermore, the SP and NW basins maintain SSTs above 28°C throughout the period. In SON (Figure \ref{fig2}  Panel d), SSTs across all basins remain elevated before the event, with SP and NW exceeding 28.5°C. Post-event cooling is modest in magnitude and mostly limited to SI, which shows a shallow decline followed by a recovery.  A general synthesis of the seasonal SST responses across basins reveals a consistent post-cyclone cooling pattern. In most cases, the lowest SST is reached between 3 and 7 days following the cyclone event, reflecting the prompt adjustment of the upper ocean to cyclone-induced vertical mixing. The NWP basin shows the least pronounced SST variations associated with cyclone activity, whereas the  SI and SP basins consistently exhibit greater fluctuations across all seasons.

\begin{figure}[ht]
\centering
\includegraphics[width=\linewidth]{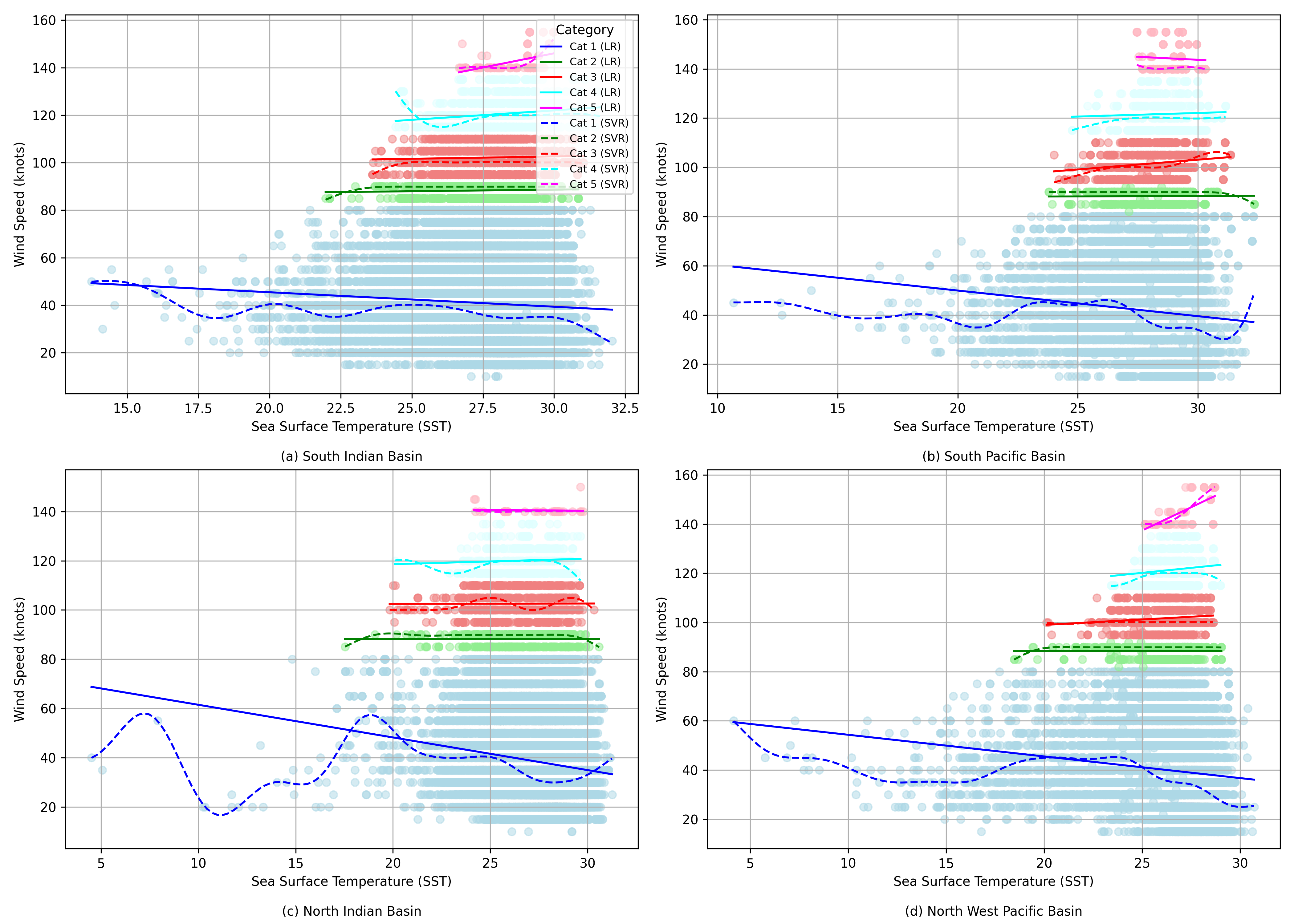}
\caption{Relationship between cyclone wind intensity and SST across four basins: Linear regression (solid line) vs. SVR (dashed line). The four subplots, labelled (a), (b), (c), and (d) in sequence, respectively, represent the South Indian Ocean Basin, Northwest Pacific Ocean Basin, South Pacific Ocean Basin and North Indian Ocean Basin.}
\label{fig3}
\end{figure}

We designed experiments to investigate the relationship between SST and tropical cyclone intensity. We employed two modelling approaches to fit the data, including linear regression and  Support Vector Regression (SVR) \cite{smola2004tutorial} as a machine learning method, with results in Figure \ref{fig3}.  In the South Indian Ocean (Figure \ref{fig3} Panel a), wind speed tends to increase with SST in the range of 25–30°C, particularly for higher-category cyclones (Categories 3–5). Both SVR and linear regression indicate a positive association within this temperature band, though SVR captures a more nuanced upward curvature at extreme SST values. Below ~25°C, the relationship flattens considerably, suggesting a lower likelihood of cyclone intensification under cooler surface conditions.  In the Northwest Pacific Ocean (Panel b), the pattern is broadly similar, with most intense cyclones clustered between 27–30°C SST. However, SVR fits indicate a subtle drop-off in wind speed at SSTs above ~30°C. The linear model fails to capture this non-linearity, emphasising the advantage of SVR for representing upper-end variability. In the South Pacific Ocean (panel c), SSTs below 25°C are more common, and high-intensity cyclones are correspondingly less frequent. The fitted curves for both methods are relatively flat until SSTs exceed ~26°C, beyond which wind speeds increase sharply, particularly for Categories 4 and 5. This threshold-like behaviour suggests a possible SST tipping point necessary for major cyclone development in this basin. The North Indian Ocean (Panel d) displays the most defined increase in wind speed with rising SST. Both regression models show a steep and continuous positive slope above ~26°C, and Category 5 cyclones are confined almost exclusively to SSTs above 28°C.

 Overall, the results demonstrate a consistent positive correlation between SST and cyclone intensity across all basins, although the strength and shape of the relationship vary regionally. The SVR model captures nonlinear trends more effectively than simple linear models, particularly at temperature extremes. These findings underscore the value of basin-specific modelling approaches in understanding cyclone-SST dynamics and highlight SST thresholds that could be critical for forecasting high-intensity events.

\section{Discussion}

Our findings are consistent with recent studies conducted in various oceanic regions \cite{xu2024composited,lin2024typhoon,yu2023tropical}. These aligned patterns across different regions strongly support changes in SST before and after cyclone events (Figure 7). The  SST fluctuations before and after cyclone events have predictable increases and decreases in different environments, especially for Category 5 cyclones in the South Indian Ocean and North Indian Ocean basins (Figure 8). Tropical cyclone formation requires warm ocean waters and low wind shear, since higher SSTs provide ample energy for cyclones as supported by other investigations \cite{hibbert2023changes,regibeau2024investigating}.  In addition, there may be differences and correlations in SST between different ocean basins. For example, some scholars have mentioned that the SSTS are affected by the east-west gradient and cold-warm regions \cite{seager2022persistent,conde2025observed}. The regional diversity of our dataset allows for in-depth analysis of this type of question.

We note that in terms of the number of cyclones, the previous decade (2010-2020) reported a much lower number of cyclones when compared to the past for the South Pacific and North Indian Ocean basins (Figure 4).  This could be due to the fluctuations in the decadal SST for these regions.

Our dataset has some limitations that could impact the analysis. One issue is that it includes relatively few features related to tropical cyclones, which might limit the depth of our findings. Another limitation is the time interval between cyclone records, which is set at 6 hours. This interval is quite long and may miss short-term changes in sea surface temperature and cyclone characteristics, potentially underestimating rapid variations during cyclone events. Improving these aspects in future studies could lead to more accurate results.

In future work, the framework can be extended by enabling scripts to obtain real-time data and updating the dataset automatically as data becomes available. Furthermore, the models can also be automated so that they can update their analysis as more data is available.

\section{Conclusions}

We applied two models (linear regression and SVR) to fit the data and illustrate the relationship between SSTs and tropical cyclones in different ocean basins. First, we observed that in the 15-day window preceding cyclone genesis, the SST was relatively stable or slightly increased, while after the formation of a cyclone, the SST in all ocean basins dropped significantly. This pattern of SST variation is usually more obvious in the Southern Hemisphere region (South Indian and South Pacific). Notably, we found strong positive correlations between SST and high wind speed (Category 5 Cyclones) across the South Indian Ocean and the North Indian Ocean via linear regression and SVR approaches.

The major contribution of this study is in the amalgamation of cyclone data with SST data obtained from two different sources. Furthermore, we are providing open-source code so that the data can be extended as fine-grained data points (hourly interval), and more features are available. Furthermore, SST data can be sourced from different venues and combined for uncertainty quantification. 

\section*{Code and data availability}

The code used to implement downloading the raw data, merging the raw data, generating the new dataset, and the cyclone-SST dataset can be accessed through the Github repository available at \url{https://github.com/sydney-machine-learning/cyclonedata-SST-GCM}.

\section*{Usage Notes}
The columns in each dataset are: \\
\texttt{Basin}: Basin where TCs occur; \\
\texttt{No\_cyclone}: The label of each TCs; \\
\texttt{Time}: from JTWC, the time that the TCs got recorded; \\
\texttt{Lat}: Latitude of the TCs record location; \\
\texttt{Lon}: Longitude of the TCs record location; \\
\texttt{lat\_tenth}: The numerical of latitude; \\
\texttt{lon\_tenth}: The numerical of longitude;\\
\texttt{SST}: from OI-SST, SST during the passage of a TCs; \\
\texttt{AVG\_SST}: Mean SST for the corresponding month and position; \\
\texttt{RSST}: The difference between SST and AVG\_SST;\\
\texttt{Speed(knots)}: from JTWC, wind intensity at the TC centre; \\
\texttt{Month}: Tropical cyclone in which month; \\
\texttt{Season}: Season of the TCs (There are four separate seasons. The initial letter of each month within a season serves as the variable name; for instance, MAM denotes March to May); \\
\texttt{Saffir-Simpson\_Category}: Derived from \texttt{Speed(knots)}, it represents a classification of wind force levels, with levels from 1 to 5 representing increasing wind force; \\
\texttt{SST\_before\_i}: from OI-SST, SST at the location of TCs before i days (A total of 15 days are obtained, i from 1 to 15); \\
\texttt{SST\_after\_i}: from OI-SST, SST at the location of TCs after $i$ days; i.e. a total of 15 days in the data sample given.

\section*{Appendix}



\subsection*{Tutorial: Data processing and visualisation}

\begin{figure}[htbp]
  \centering
  \includegraphics[width=0.65\linewidth]{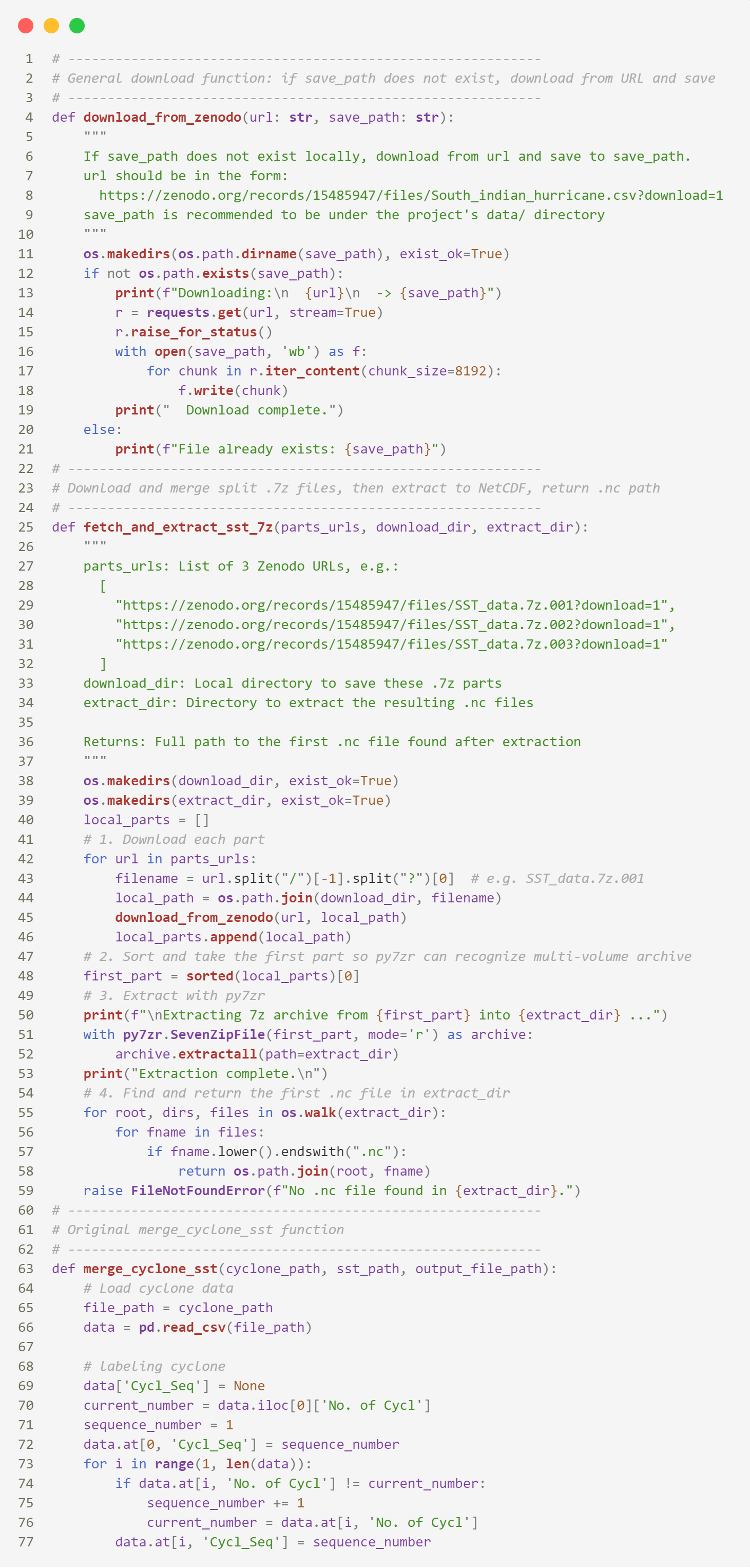}
  \caption{Python code for data downloading and dataset merging. Access from the link: \url{https://github.com/sydney-machine-learning/cyclonedata-SST-GCM/blob/main/cyclone_merge_script.py}}
  \label{code1}
\end{figure}

\begin{figure}[htbp]
\centering
\includegraphics[width=0.78\linewidth]{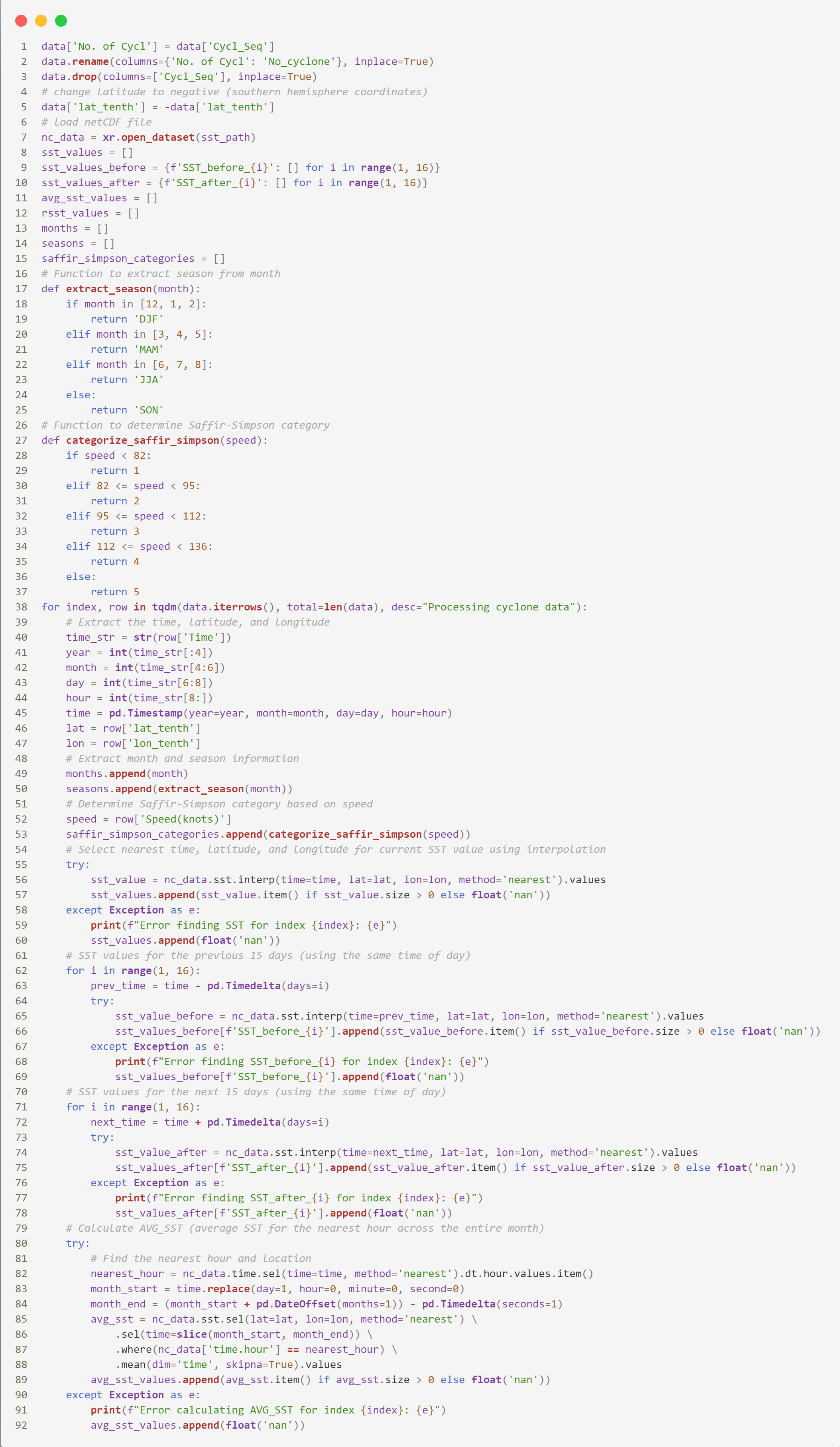}
\caption{Python code for data downloading and dataset merging (Continued from Figure \ref{code1}). Access from the link: \url{https://github.com/sydney-machine-learning/cyclonedata-SST-GCM/blob/main/cyclone_merge_script.py}}
\label{code2}
\end{figure}

\begin{figure}[htbp]
\centering
\includegraphics[width=0.8\linewidth]{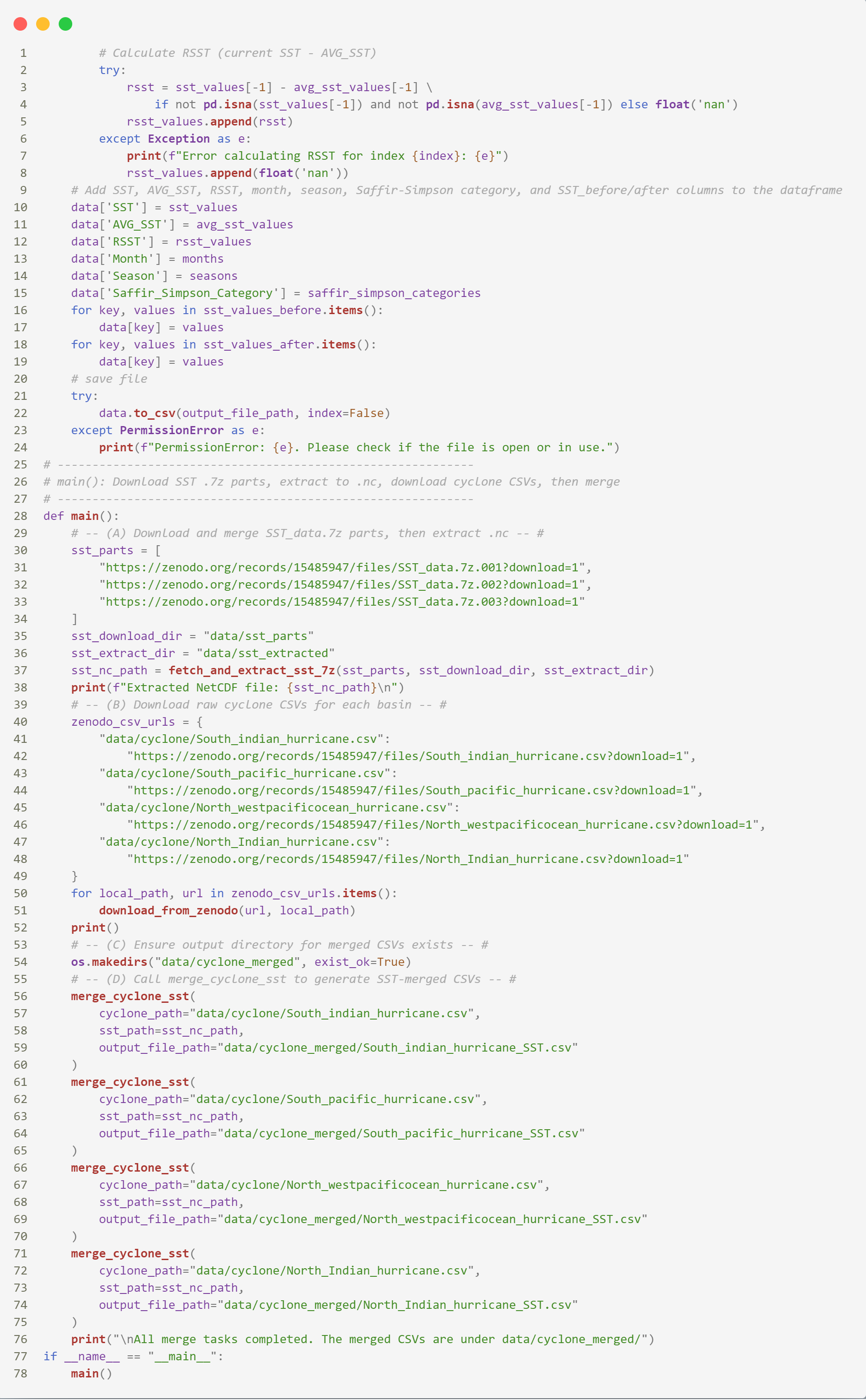}
\caption{Python code for data downloading and dataset merging (Continued from Figure \ref{code2}). Access from the link: \url{https://github.com/sydney-machine-learning/cyclonedata-SST-GCM/blob/main/cyclone_merge_script.py}}
\label{code3}
\end{figure}

This guide shows how to use the two Python scripts—\texttt{cyclone\_merge\_script.py} and \texttt{plot\_figures.py}—to reproduce all data and figures from the paper. These two Python files are in our GitHub. We present the code for \texttt{cyclone\_merge\_script.py} in Figure \ref{code1}, \ref{code2} and \ref{code3}.

\begin{itemize}
  \item Create a folder on your computer where you want to work, for example:
  \\
  \verb|projects/cyclone/|
  \item Download (or copy) the two scripts into that folder:
  \begin{itemize}
    \item \texttt{cyclone\_merge\_script.py}
    \item \texttt{plot\_figures.py}
  \end{itemize}
  \item Open a terminal and change directory into it:
  \\
  \verb|projects/cyclone/|
  \item Ensure you have Python 3.x and the required libraries installed:
   \item \texttt{pip install requests py7zr pandas xarray}\\
      \texttt{tqdm matplotlib basemap scipy pygam scikit-learn}
\end{itemize}

\subsection*{2. Step 1: Download and Prepare Data}

In your working folder, run:
\begin{itemize}
  \item \texttt{python cyclone\_merge\_script.py}
\end{itemize}

This script will:
\begin{itemize}
  \item Download three split SST archives from Zenodo into \texttt{data/sst\_parts/}. 
  \item Merge and extract them to produce \texttt{data/sst\_extracted/SST\_data.nc}.
  \item Download four raw cyclone CSVs into \texttt{data/cyclone/}.
  \item Generate four merged CSVs (with SST) in \texttt{data/cyclone\_merged/}.
\end{itemize}

After it finishes, you will see:
\begin{verbatim}
projects/cyclone/
  data/
    sst_parts/
    sst_extracted/SST_data.nc
    cyclone/            (four raw CSVs)
    cyclone_merged/     (four merged CSVs)
  cyclone_merge_script.py
  plot_figures.py
\end{verbatim}

\subsection*{3. Step 2: Generate Figures}

\begin{figure}[htbp]
\centering
\includegraphics[width=0.8\linewidth]{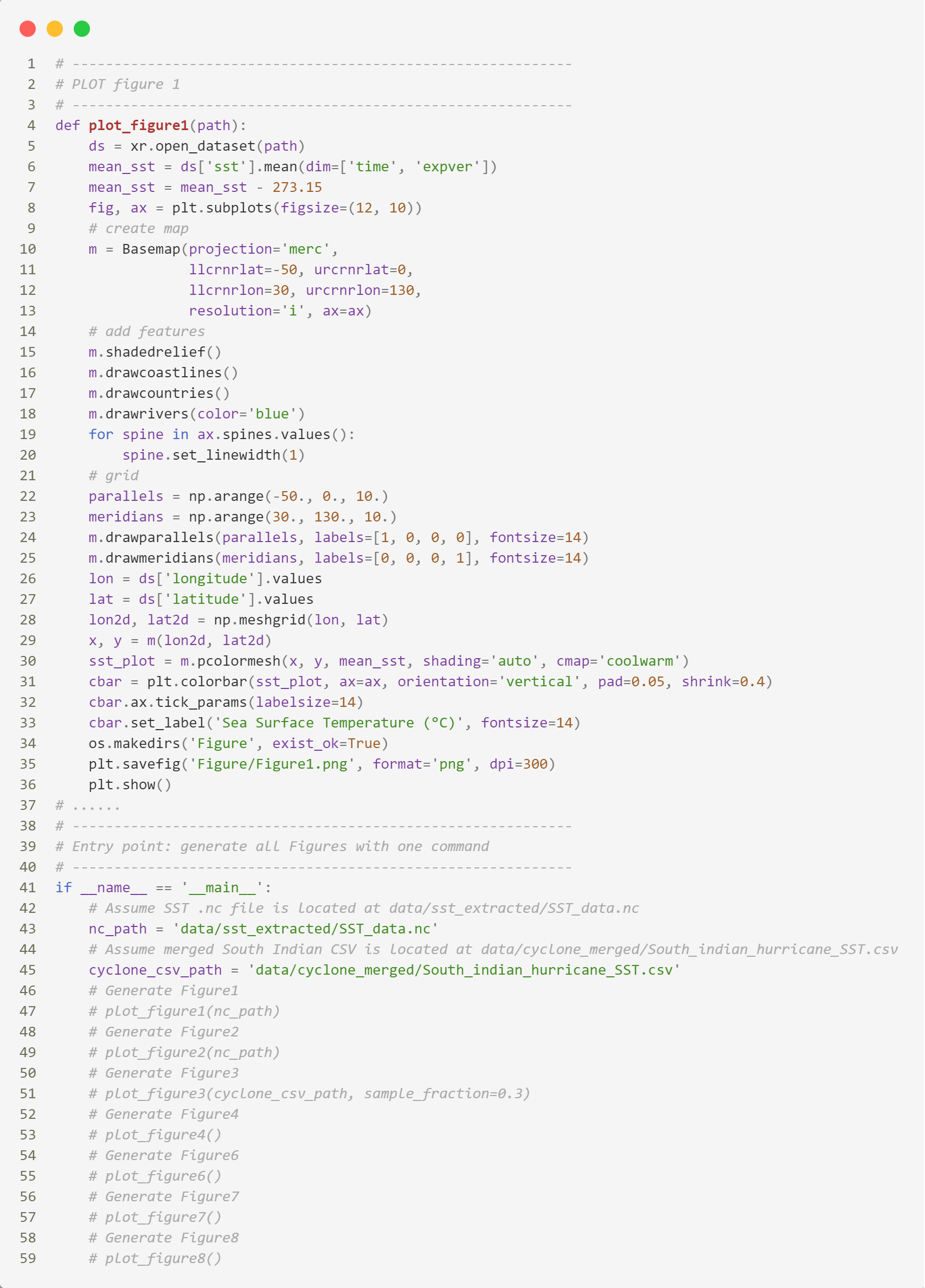}
\caption{Python code for plotting the figures. Access from the link: \url{https://github.com/sydney-machine-learning/cyclonedata-SST-GCM/blob/main/plot_figures.py}}
\label{code4}
\end{figure}

The code for plotting the figures is shown in Figure \ref{code4}. We show the generated code of Figure \ref{fig_si_avgsst} and the main function of the entire Python file
Open \texttt{plot\_figures.py} in a text editor. Near the bottom, you will find main function.

Uncomment the functions you want, then run \texttt{plot\_figures.py} to generate figures. The generated figures will be saved in the Figure folder at the root directory.

Then run:
\begin{verbatim}
python plot_figures.py
\end{verbatim}

Each \texttt{plot\_figureX(...)} call will:
\begin{itemize}
  \item Read the appropriate NetCDF or merged CSV from \texttt{data/}.
  \item Create the figure and save it into a new \texttt{Figure/} folder.
\end{itemize}

After running, you will see:
\begin{verbatim}
projects/cyclone/
  Figure/
    Figure1.png
    Figure2.png
    Figure3.png
    ... (others as uncommented)
  data/
    ... (as before)
  cyclone_merge_script.py
  plot_figures.py
\end{verbatim}

\section*{Acknowledgements}  

 We thank Nandini Ramesh, Albert Demskoy, and Joshua Simmons.

\bibliography{sample}

\end{document}